\documentclass[12pt]{article}
\usepackage{color,graphicx}
\usepackage{hfstyle}

\begin{document}
\title{Individual-based lattice model for spatial spread of
epidemics}
\author{Henryk Fuk\'s$^{1,4}$ and Anna T. Lawniczak$^{2,3,4}$
      \oneaddress{
         $^1$ Department of Mathematics\\ Brock University\\
         St. Catharines ON L2S 3A1, Canada \\
         {~}\\
         $^2$ Department of Mathematics and
         Statistics \\
         $^3$ Guelph-Waterloo Institute of Physics\\
         University of Guelph, \\
         Guelph, Ontario N1G 2W1, Canada\\
         {~} \\
         $^4$ The Fields Institute for Research\\
         in Mathematical Sciences\\
         Toronto, Ontario M5T 3J1, Canada\\
         {~}\\
         \email{hfuks@brocku.ca,  alawnicz@fields.utoronto.ca}
       }
   }
\Abstract{ We present a lattice gas cellular automaton (LGCA)
to study spatial and temporal dynamics of  an epidemic of SIR
(susceptible-infected-removed)  type.
The automaton is fully
discrete, i.e. space, time and number of individuals are discrete variables.
The  automaton can be applied to study spread of epidemics in
both human and animal populations. We investigate
effects of spatial inhomogeneities in initial distribution of  infected and vaccinated
populations on the dynamics of epidemic of SIR type. We discuss
vaccination strategies which
differ only in spatial distribution of vaccinated individuals.  Also, we
derive an  approximate, mean-field type description of the
automaton, and discuss differences between the mean-field dynamics and the results
of LGCA simulation.}
\maketitle

\section{Introduction}

Since the publication of Kermack and McKendrick epidemic model \cite
{kermack27} in 1927, mathematical epidemiology developed an extensive body
of literature. A vast majority of proposed models, to study dynamics of
epidemics, is based on ordinary differential equations. These models assume
that the population mixing is strong, hence concentrations of effected types
of population (for example, susceptible, infected, removed) are spatially
homogeneous. Thus, these models neglect spatial aspects of the epidemic
process. Models which rely on partial differential equations, such as \cite
{murray86}, abandon the assumption of homogeneous mixing and allow to study
the geographical spread of epidemics, yet they still pose some serious
problems. They treat the population as continuous entity, and neglect the
fact that populations are composed of single interacting individuals. This
can lead to very unrealistic results, such as, for example, endemic patterns
relaying on very small densities of individuals, named in \cite{mollison91}
``atto-foxes'' and ``nano-hawks''.

Models based on interacting particle systems eliminate these shortcomings of
traditional methodologies. They treat individuals in biological populations
as discrete entities, allow to introduce local stochasticity, and are well
suited for computer simulations. Several models of this type have been
suggested in a recent decade, including stochastic interacting particle
models \cite{durret94}, and models based on cellular automata or coupled map
lattices \cite{schon93,bc93,ahmed98,duryea99,benyo2000}.

In this article, we present lattice gas cellular automaton (LGCA) for SIR
(susceptible-infected-removed) type of an epidemic. Our aim is to
demonstrate that the LGCA approach provides an interesting and potentially
fruitful alternative to other methods. In real biological populations, both
animal and human, contact processes among infectious and susceptible
individuals and their movement play a vital role in spreads of epidemics of
an infectious disease. Infectious diseases spread because infectious and
susceptible individuals mix together. They move, meet each other and through
a contact process they transmit an infection. Hence, among other factors the
spread of infectious diseases strongly depends on patterns of mobility in
populations. In LGCA, population mixing arises directly from motion of
individuals and susceptible individuals can become infected only if they
meet infectious individuals. Additionally, a LGCA allows to investigate
effects of spatial inhomogeneities in population concentrations on the
dynamics of epidemic processes and vaccination strategies. We will
demonstrate examples of such effects. In this article we do not aim to
construct a LGCA of a particular disease but rather we want to study generic
features of LGCA methodology and its suitability in the context of
epidemiology. To illustrate the ideas and discuss further possible
developments we selected SIR (susceptible-infected-removed) epidemic type
for which we constructed a LGCA. We derived approximate mean-field type
description of the automaton and simulated and analyzed the automaton
dynamics. We compared the predictions obtained from the mean-field
description with those obtained from the automaton simulations.
Additionally, we used automaton to study various vaccination strategies.

\section{Individual-based SIR model on a lattice}

We will construct a LGCA for an epidemic of SIR type, for which we can
assume that a population consists of three types of moving and interacting
individuals, of type $\tau \in \{S,I,R\}$, susceptible, infected, and
recovered. The proposed automaton is a special case a
lattice gas cellular automaton for reaction-diffusion systems,
described in detail in \cite{boon96,law99}.

\begin{figure}[tb]
\begin{center}
\includegraphics{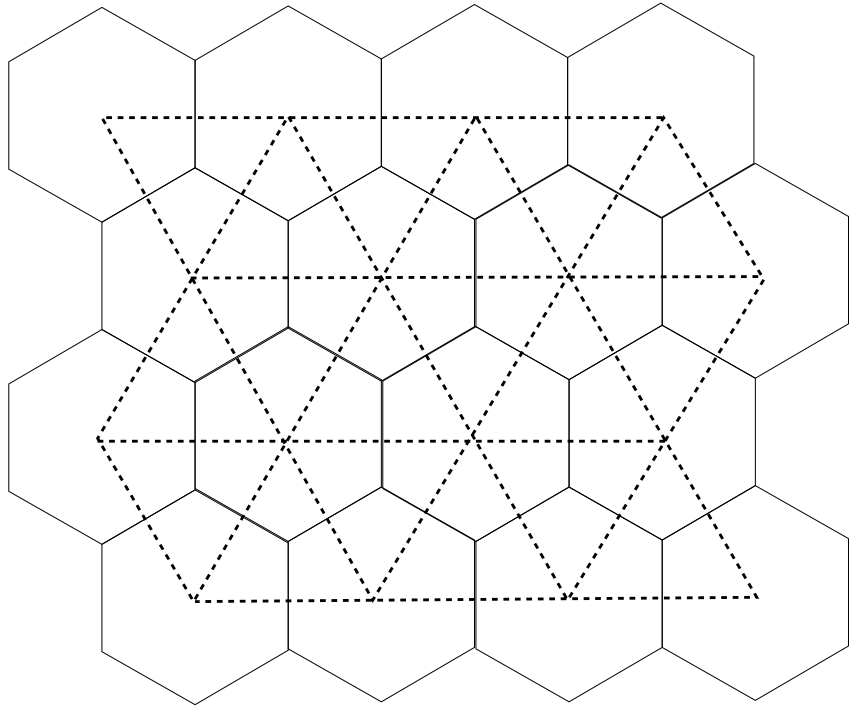}
\end{center}
\caption{The array of hexagonal cells with a dashed lattice superimposed on
it. The lattice depicts the channels through which the individuals are
transported from cell to cell. The figure is reproduced from \cite{vl99}.}
\label{hex}
\end{figure}

We tile the physical space, in which an epidemic takes place, by regular
hexagonal cells $h(\mathbf{r)}$ with centers at discrete space variables $%
\mathbf{r}$. We chose the hexagonal cells in order to avoid spurious
invariants in the dynamics of the automaton, also known as parity problem
\cite{boon96,law99}. However, other types of tiling, such as square cells,
can be used as well. We assume that the distance between centers of adjacent
cells is $1.$ Let
\begin{equation}
\mbox{\boldmath$ c$}_{i}=<\cos (i-1)\pi /3,\sin (i-1)\pi /3>
\end{equation}
be a unit vector, for each $i=1,...,6.$ If we connect the center of every
hexagon $h(\mbox{\boldmath$ r$})$ with the centers of the neighbouring
hexagons $\mbox{\boldmath$ r$}+\mbox{\boldmath$ c$}_{i}$, where $i=1,...,6,$
then we obtain a hexagonal lattice structure $\mathcal{L}_{h}$, with the
lattice coordination number $m=6.$ In the case of square cells the lattice
coordination number would be $m=4.$ Figure \ref{hex} (reproduced
from \cite{vl99}) shows that the
centers of the hexagons $h(\mbox{\boldmath$ r$})$ become the nodes of the
hexagonal lattice $\mathcal{L}_{h}.$ The cells represent some area of the
physical space in which the individuals mixed, interact and can move from
one place (cell) to another. The individuals can move among the cells only
along the channels corresponding to edges of the lattice $\mathcal{L}_{h}.$
Except where confusion might arise, we will identify a cell $h(%
\mbox{\boldmath$ r$})$ with its center $\mbox{\boldmath$ r$}$, hence with
the node $\mbox{\boldmath$ r$}$ of the lattice $\mathcal{L}_{h}.$ Hence, we
can say that the individuals reside at the nodes of the lattice and move
along its edges (channels). While extensions to arbitrary boundary
conditions are straightforward, we limit our considerations to periodic
boundary conditions.

At the initial time, $k=0,$ particles representing individuals $S,$ $I$ and $%
R$ are distributed randomly and independently, among the cells $h(%
\mbox{\boldmath$
r$}),$ $\mbox{\boldmath$ r$}\in \mathcal{L}_{h},$ according to probabilities
given by their initial concentrations. The initial distribution of particles
is such that there are at most six particles, regardless of the individual
type, per cell $h(\mbox{\boldmath$ r$})$ (four when the cells are squares).
The LGCA discrete dynamics is constructed in such a way that the total number
of individuals at a given node is restricted to stay between $0$ and $m$ ($%
m=6$ or $4$, respectively, for hexagonal or square lattice) per each cell
during the time evolution of the automaton. Since a single lattice node
represents a small region of space, this means that $m$ can be interpreted
as the carrying capacity of that region. In other words, the number of
individuals that this region can support cannot exceed $m$.

The time evolution of the automaton takes place at discrete time steps. At
each time step $k$, an evolution operator $\mathcal{E}$ is applied,
simultaneously\ and independently of the past, to all lattice nodes. The
evolution operator $\mathcal{E}$ governs the dynamics of the automaton,
which arises from sequential applications of the three basic operations
\textit{contact}~$\mathcal{C}$, \textit{randomization }$\mathcal{R}$ and
\textit{propagation} $\mathcal{P}$. Hence, the evolution operator can be
written in terms of these operations as the superposition
\begin{equation}
\mathcal{E}=\mathcal{P\circ R\circ C}.
\end{equation}
Each of the operations $\mathcal{C}$, $\mathcal{R}$ and $\mathcal{P}$
captures some aspects of the epidemic process and their actions are
described as follows.

\begin{enumerate}
\item  As a result of an application of the \textit{contact operation} $%
\mathcal{C}$, individuals can change their type, meaning that susceptible
individuals can become infected, and infected individuals can recover. More
precisely, each susceptible individual at a node $\mathbf{r}$, independently
of other individuals, can become infected with probability $1-(1-r)^{n_{I}}$%
, where $n_{I}$ is a number of infected individuals at the node $\mathbf{r}$%
, and $r\in \lbrack 0,1]$. Similarly, each infected individual at the node $%
\mathbf{r}$, independently of other individuals, can recover with
probability $a$, where $a\in \lbrack 0,1]$.

\item  As a result of using the \textit{randomization operation} $\mathcal{R}
$, applied at each node independently of the other nodes, a population of
individuals residing at a node $\mathbf{r}$ is randomly redistributed among
edges/channels originating from the node $\mathbf{r}$. Through the selected
channels, allowing at most one individual per channel, individuals will move
in the propagation step from the node $\mathbf{r}$ to the neighbouring
nodes. The process of redistribution of individuals is purely probabilistic
one and it contributes to modeling the mixing process of individuals.

\item  In the \textit{propagation step,} governed by the operator $\mathcal{P%
}$, individuals simultaneously move from their nodes to the neighbouring
ones through the channels assigned to them in the randomization step. The
movement of individuals is purely deterministic in the propagation step.
\end{enumerate}

The rationale behind the probability of becoming infected in the first step being $%
1-(1-r)^{n_{I}}$ can be explained as follows. We assume that, at a given
node, a susceptible individual contacts all infected individuals at that
node and that all infected individuals are infectious one. If the
probability of infection per contact is $r$, then the probability of \textit{%
not} getting infected after contacting each of $n_{I}$ infected individuals
is $(1-r)^{n_{I}}$. The probability of getting infected, therefore, is $%
1-(1-r)^{n_{I}}$.

Note that the mechanism of contracting an infection described here implies that
the incubation
period is short enough to be negligible, meaning that a susceptible who
contracts the disease at time $k$ becomes infectious immediately, and can
infect others at time $k+1$. While this is a convenient assumption, it will
have to be relaxed in more realistic models, or in models tailored for a
specific disease. Also, note that the duration of the disease (number of time
steps from contracting an infection to recovery) has geometrical
distribution with a mean value of $1/a$ time steps. Again, in realistic
disease-specific models a different distribution will have to be adopted,
depending on the particular disease. As we said earlier the spread of
infectious diseases strongly depends on mixing patterns in the populations.
In the LGCA that is being presented here, mixing is of diffusive type. It
arises from randomization of directions of motion of individuals in the
randomization step $\mathcal{R}$ and the movement of individuals in the
propagation step $\mathcal{P}$. In the more realistic models the movement of
individuals will have to be appropriately modified and additionally, we will
have to include inflow, outflow, birth and death processes.

\section{Mean-field approximation}

In order to gain some insight into the dynamics of the LGCA defined in the
previous section, we will now proceed to construct approximate equations
describing the automaton dynamics. Using the formalism and methodology
introduced in \cite{law99}, it is possible to write compact microdynamical
equations corresponding to the evolution operator $\mathcal{E}$. Moreover,
by making appropriate approximations, it is possible to derive the LGCA
discrete lattice-Boltzmann equations and the corresponding partial
differential equations describing the automaton macroscopic dynamics. Since
rigorous derivation of these equations is rather involved, we refer the
interested reader to \cite{law99}. We take here a less rigorous, but more
intuitive approach.

Let us assume that the total number of nodes in the lattice $\mathcal{L}_{h}$
is $N$, and at time $k$ there are $N_{S}(k)$ susceptible, $N_{I}(k)$
infected, and $N_{R}(k)$ recovered individuals on the entire lattice. Since
we do not allow for inflow, outflow, birth and death processes, the total
population, denoted by $N_{tot},$ remains constant in time. Define $n_{\tau
}(\mathbf{r},k)$ to be a number of individuals of type $\tau $ at a node $%
\mathbf{r}$ at time $k$. If we assume that the automaton dynamics is
spatially homogeneous (well stirred), then the expected value of $n_{\tau }(%
\mathbf{r},k)$ is independent of $\mathbf{r}$ and equals
\begin{equation}
\langle n_{\tau }(\mathbf{r},k)\rangle =\frac{N_{\tau }(k)}{N}.
\end{equation}
Under the same assumptions, regardless at which cell the susceptible
individual is located, the expected number of infected individuals at his
cell is $N_{I}/N$, thus the probability that he becomes infected is
\begin{equation}
1-(1-r)^{N_{I}(k)/N}.
\end{equation}
Hence, the expected number of susceptible individuals who become infected in
a single time step is equal to
\begin{equation}
N_{S}(k)[1-(1-r)^{N_{I}(k)/N}].
\end{equation}
Similarly, the expected number of individuals who become recovered in a
single time step is $aN_{I}(k)$. When the population is well stirred this
yields that the expected number of individuals of each type $\tau \in
\{S,I,R\}$ at time $k+1$ is
\begin{eqnarray}
N_{S}(k+1) &=&N_{S}(k)(1-r)^{N_{I}(k)/N},  \nonumber \\
N_{I}(k+1) &=&N_{I}(k)+N_{S}(k)[1-(1-r)^{N_{I}(k)/N}]-aN_{I}(k), \\
N_{R}(k+1) &=&N_{R}(k)+aN_{I}(k).  \nonumber
\end{eqnarray}
For small $r$, taking Taylor expansion
\begin{equation}
(1-r)^{N_{I}(k)/N}=1-\frac{rN_{I}}{N}+\frac{N_{I}(N_{I}-N)r^{2}}{2N^{2}}%
+\ldots ,
\end{equation}
and by keeping only the first two terms and defining $\rho _{\tau
}(k)=N_{\tau }(k)/N$, we obtain
\begin{eqnarray}
\rho _{S}(k+1) &=&\rho _{S}(k)-r\rho _{S}(k)\rho _{I}(k),  \nonumber
\label{mfarho} \\
\rho _{I}(k+1) &=&\rho _{I}(k)+r\rho _{S}(k)\rho _{I}(k)-a\rho _{I}(k), \\
\rho _{R}(k+1) &=&\rho _{R}(k)+a\rho _{I}(k).  \nonumber
\end{eqnarray}
The above equations are quite similar in their structure to the ordinary
differential equations obtainable from the classic Kermack-McKendrick model%
\footnote{%
The Kermack-McKendrick model \cite{kermack27} is based on integral equation
with general time-kernel for infectivity. System of ODE we are referring to
can be obtained from this integral equation as a special case.}, as
described, for example, in \cite{murray89}. The similarity lies in the fact
that the gain in the class of infected individuals occurs at a rate
proportional to the density of infectives and susceptibles, in analogy to
the mass action law in chemical kinetics. Of course, this is valid only for
small values of $r$ and under the assumption of strong mixing, which is not
always realistic.

\begin{figure}[tb]
\begin{center}
\includegraphics{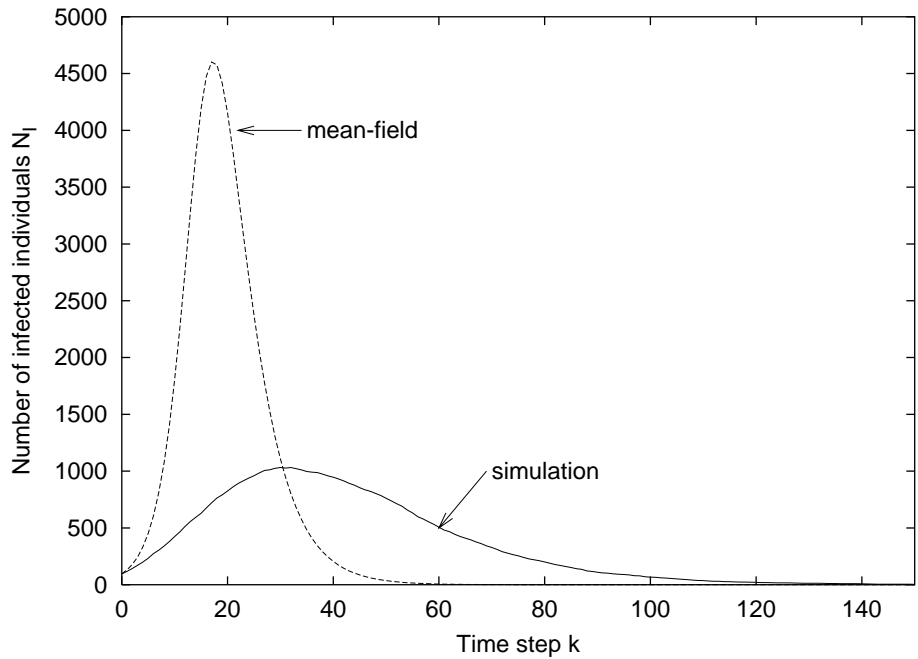}
\end{center}
\caption{Number of infected individuals as a function of time, comparison of
LGCA simulation results with the mean-field approximation. The simulation
curve represents average of 50 LGCA experiments. Each LGCA simulation has
been performed on a hexagonal lattice with $10^{4}$ nodes with $r=0.3$, $%
a=0.2$, and with $16000$ susceptibles and $100$ infected individuals
uniformly and randomly distributed on the lattice at $k=0$. \ }
\label{figureMF}
\end{figure}

In reality, the epidemic process spreads quite differently than the
mean-field approximation (\ref{mfarho}) predicts. Figure \ref{figureMF}
compares the number of infected individuals as a function of time as
observed in the LGCA simulations with the mean-field approximation (\ref
{mfarho}). The simulations have been performed on a hexagonal lattice with $%
10^{4}$ nodes, using $r=0.3$, $a=0.2$. The initial configuration consisted
of $16000$ susceptibles and $100$ infected individuals for each
simulation.
At the initial time for each simulation the individuals were randomly and
uniformly distributed on the lattice. The simulation
curve in Figure \ref{figureMF} represents average over 50 experiments. The
mean-field approximation (MFA) predicts much faster initial spread than
observed in the LGCA simulations. This can be easily explained by realizing
that the MFA assumes perfect mixing, meaning that an infected individuals
can always infect some susceptibles residing at their node. In simulations,
since the mixing is limited, it often happens that the number of infected
individuals in the vicinity of an infected individuals is larger than
average and, similarly, the number of susceptibles is smaller than average.
Or, in other words, infected individuals are often grouped together in small
regions of space, while in other spatial regions there are no infected
individuals at all. Therefore, the effective force of infection is smaller
than predicted by MFA. 
We should also note that the fraction of susceptibles who eventually became
infected is $67\%$ for simulations and $94\%$ for mean-field, meaning the
epidemic is less severe when the mixing is weaker, even though it lasts
longer, as can be seen in Figure \ref{figureMF}.


\section{Effects of spatial distribution}

In order to illustrate the importance of spatial distribution of individuals
in spread of epidemics, let us consider the following problem. Let us assume that in
some region of space, to be denoted by $A$, several cases of an infectious
disease have been reported. In order to limit the spread of the disease, we
would like to vaccinate the population. However, we have only $M$ doses of a
vaccine at our disposal, where $M$ is less than the number of susceptible
individuals in the population. The natural question arises how should these
doses be distributed in the population to minimize the severity of the
epidemic? 
\begin{figure}[tb]
\begin{center}
\includegraphics{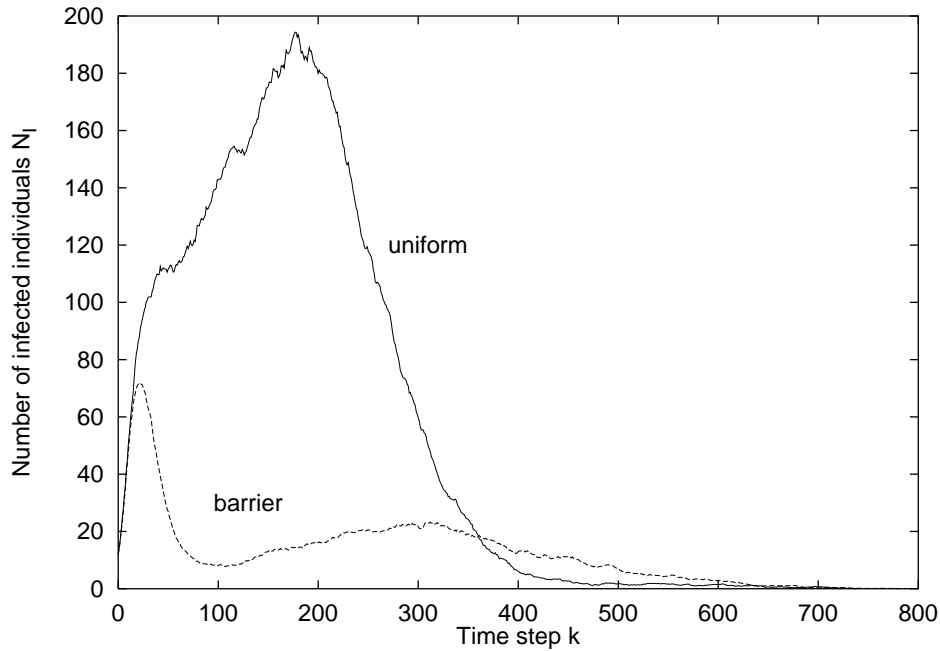}
\end{center}
\caption{Number of infected individuals as a function of time, under
``uniform'' and ``barrier'' vaccination strategies. Each simulation curve
represents average of 100 LGCA experiments. For each type of vaccination
strategy each simulation has been performed on a hexagonal lattice with $%
10^{4}$ nodes with $r=0.3$, $a=0.2$, and with $16000$ susceptibles uniformly
and randomly distributed on the lattice at $k=0$, and $10$ infectives
uniformly and randomly distributed in the circle of radius 20 on the lattice
at $k=0$. }
\label{patdat}
\end{figure}
To simplify the problem, we assume that the vaccine is immediately
effective, therefore vaccinated individuals become members of the \textit{%
recovered} group immediately after vaccination. Additionally, we assume that
the region $A$ is a circle of radius $20$ on a hexagonal lattice of $10^{4}$
nodes.

\begin{figure}[p]
\begin{center}
\begin{tabular}{rcccl}
(a) & \scalebox{1.0}[0.66]{\includegraphics{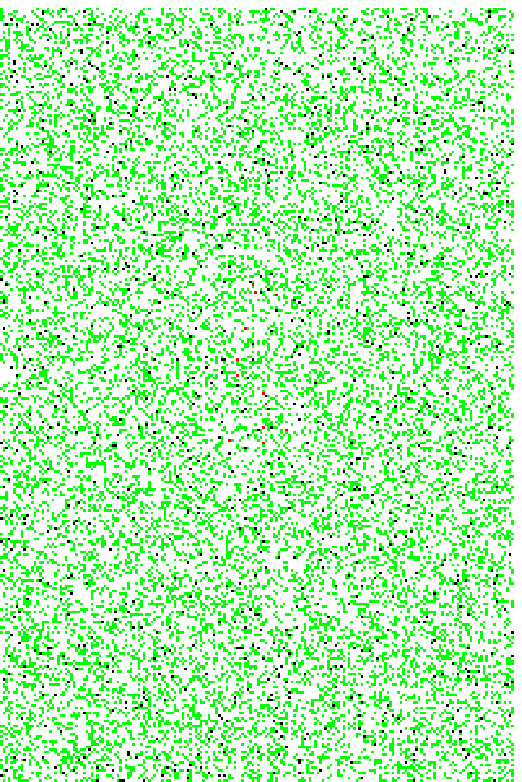}} & $k=0$ & %
\scalebox{1.0}[0.66]{\includegraphics{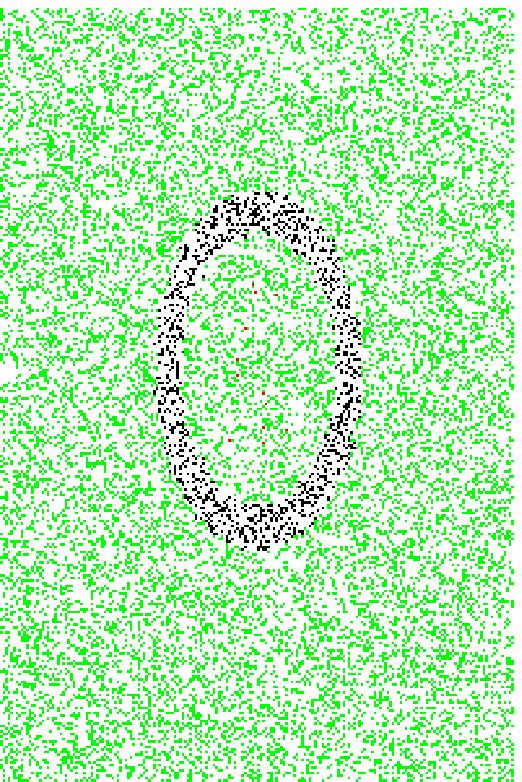}} & (e) \\
(b) & \scalebox{1.0}[0.66]{\includegraphics{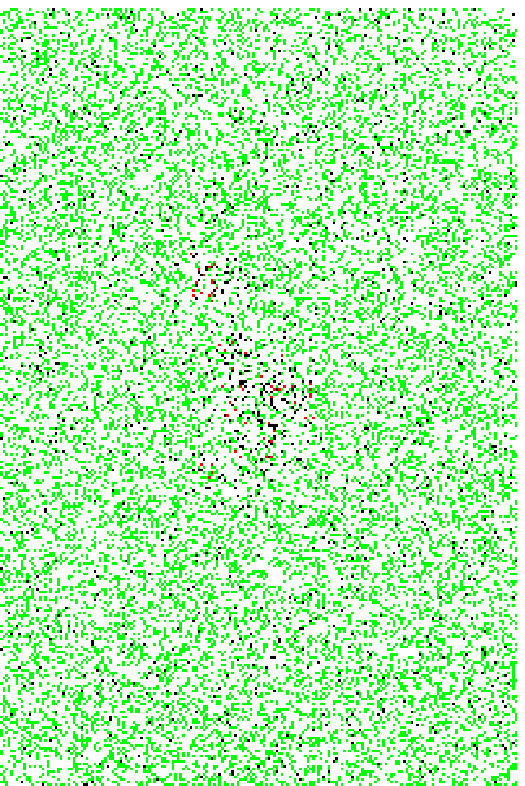}} & $k=20$ & %
\scalebox{1.0}[0.66]{\includegraphics{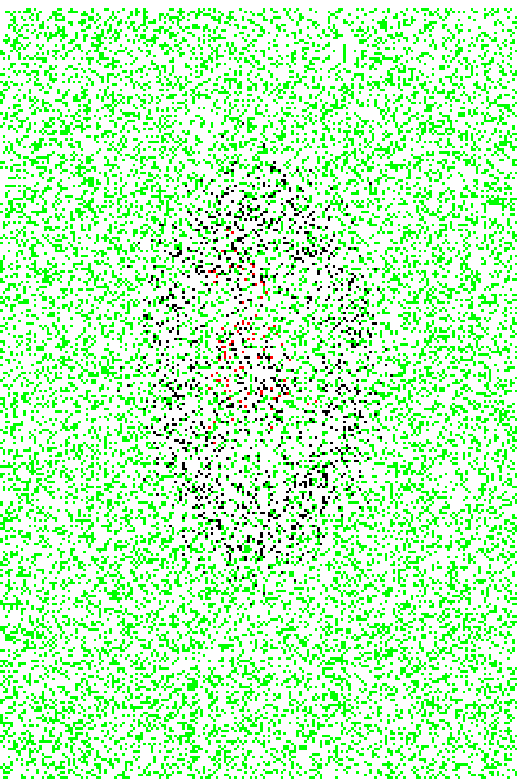}} & (f) \\
(c) & \scalebox{1.0}[0.66]{\includegraphics{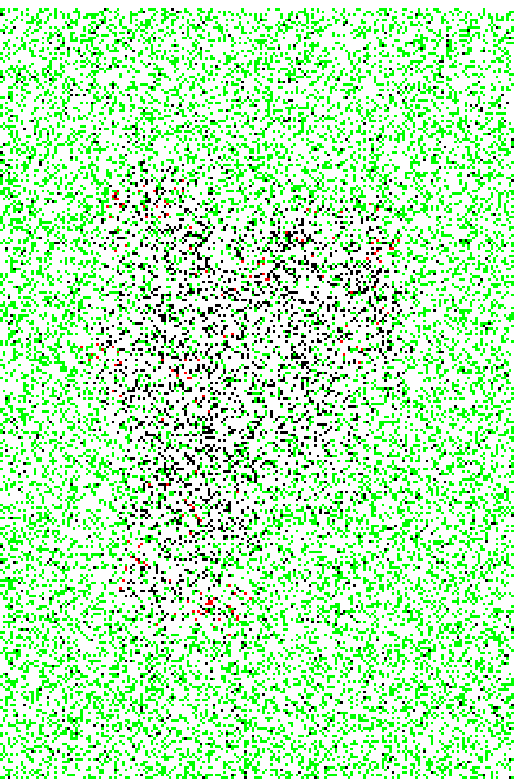}} & $k=100$ & %
\scalebox{1.0}[0.66]{\includegraphics{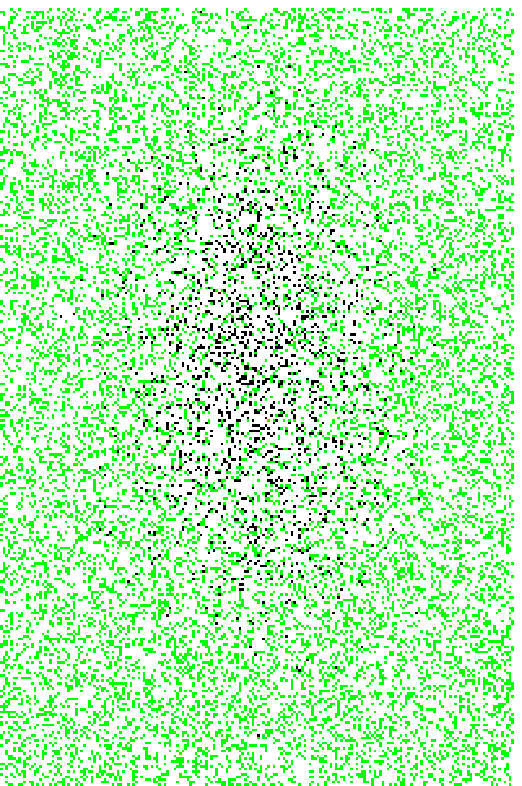}} & (g) \\
(d) & \scalebox{1.0}[0.66]{\includegraphics{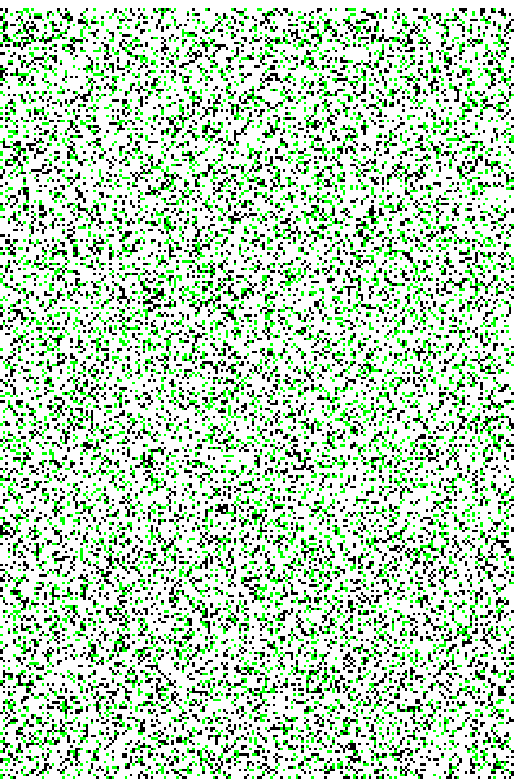}} & $k=1000$ & %
\scalebox{1.0}[0.66]{\includegraphics{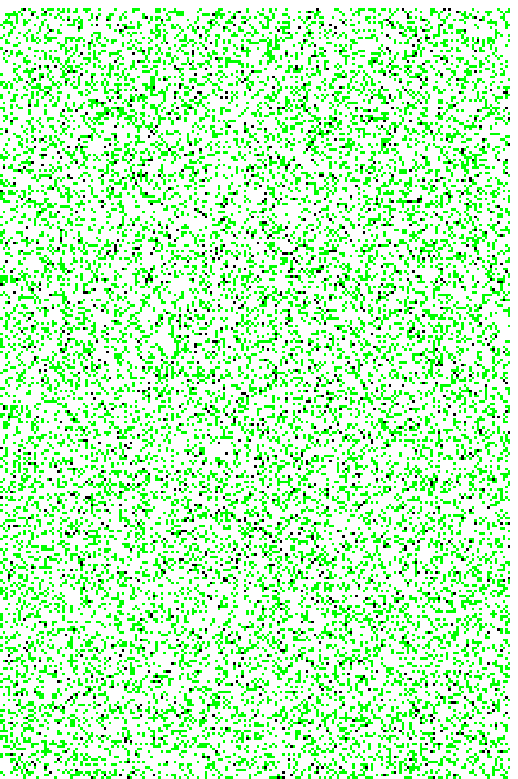}} & (h)
\end{tabular}
\end{center}
\caption{Snapshots of LGCA dynamics evolution under ``uniform'' (left) and
``barrier'' (right) vaccination strategies.}
\label{figurepatt}
\end{figure}

We compare two alternative vaccination strategies. In the first one, to be
referred to as a ``uniform strategy'', we vaccinate $M$ individuals selected
from the entire population of susceptibles randomly and independently of
each other. In the second strategy, to be called a ``barrier strategy'', we
vaccinate all individuals in a ring surrounding the region $A$. The
thickness of the ring is selected in such a way that the total number of
vaccinated individuals to be $M$. In all subsequent discussions we assume $%
M=1000$. Figures \ref{figurepatt}a and \ref{figurepatt}e (page~13)
show initial distribution of individuals in both these
strategies, ``uniform'' and ``barrier'', respectively. Note that the number
of individuals in $S$, $I$, and $R$ classes is the same in both cases, so
that the spatial distribution of individuals is the \textit{only} difference
between configuration in Figures \ref{figurepatt}a and \ref{figurepatt}e.
The color coding used is green for $S$, red for $I$, and black for
$R$. Consecutive snapshots of the LGCA dynamics in Fig. \ref{figurepatt} reveal
that the time evolution of these two initial configurations is quite
different. One immediately notices that the final configuration in the case
of the ``uniform'' vaccination strategy, after $1000$ time steps (Fig. \ref
{figurepatt}d), contains many more blue pixels than the corresponding final
configuration of the ``barrier'' vaccination strategy (Fig. \ref{figurepatt}%
h). This means that many more susceptible individuals contracted the disease
in the course of the epidemic when the ``uniform'' vaccination strategy was
used than in the case of the ``barrier'' vaccination strategy. To be more
precise, out of initial $16000$ susceptible individuals out of whom $1000$
have been vaccinated at the
start of epidemic, $52\%$ became infected when the vaccination was
``uniform'', and only $12\%$ when the ``barrier'' vaccination strategy was
employed. This can be seen from Figure \ref{patdat}, where the number of
infected individuals is plotted as a function of time step, averaged over $%
100$ LGCA experiments.

\section{Barrier's permeability and severity of the epidemics}

\begin{figure}[t]
\begin{center}
\includegraphics{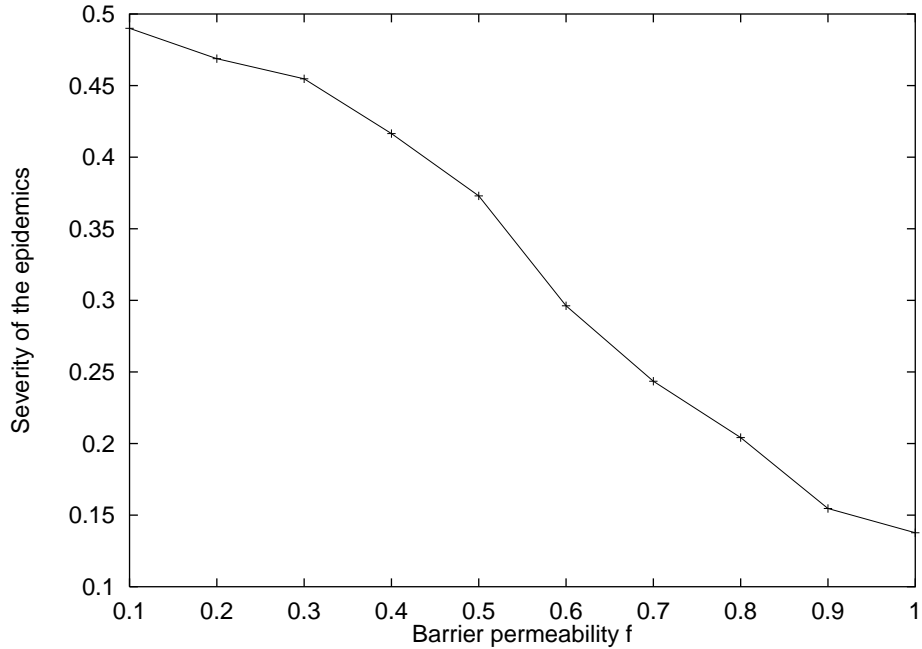}
\end{center}
\caption{Average severity of epidemics of an infectious disease $%
\langle N_{R}(\infty )-N_{R}(0)/N_{tot} \rangle$, calculated as average over severities of
outbreaks of epidemics of the disease, as a function of barrier permeability
$f$. For each $f$ the average $\langle \cdot \rangle$ has been taken over $100$ LGCA
simulations of outbreaks of epidemics, each starting with
identical initial configuration, and $r=0.3$ and $a=0.2$.}
\label{severity}
\end{figure}
Since the barrier appears to be a very effective strategy, one could ask how
permeability of the barrier influences the dynamics of the epidemics. In
what follows, we will vary permeability of the barrier by vaccinating not
all susceptibles in the ring surrounding the region $A$, but only
a fraction $f$ of randomly selected susceptibles in the ring.
Thus, if $f=0$, there is no single vaccinated individual in the ring, while $%
f=1$ corresponds to the ``barrier'' described in the previous section, where
all susceptibles in the ring are vaccinated.

We define the \textit{severity of an outbreak of an epidemic
of an infectious disease} as $(N_{R}(\infty )-N_{R}(0))/N_{tot}$,
where $N_{R}(\infty )$ denotes a total number of removed when a total number
of infected in LGCA simulation reaches a steady state and $N_{R}(0)$ denotes
the total number of removed (vaccinated individuals) at the start of an
epidemic (in an initial configuration in our case). Since in our model we do
not allow for birth, death and migration of individuals then the number of
removed will remain constant from the moment when the number of infected
reaches zero. Let $N_{S}(\infty )$ be a total number of susceptibles at
the end of an outbreak of the epidemic. Since $N_{tot}=N_{S}(0)+N_{I}(0)+N_{R}(0)=N_{S}(\infty )+N_{R}(\infty )$,
then $N_{R}(\infty )-N_{R}(0)=N_{S}(0)+N_{I}(0)-N_{S}(\infty )$ gives us the total
number of susceptibles who got infected in the course of the epidemic.
Hence, the \textit{average severity of an epidemic of infectious
disease} can be measured by taking the average $\langle(N_{R}(\infty
)-N_{R}(0))/N_{tot}\rangle$ of the severities of outbreaks of specific epidemics
of the infectious disease. It tells us on average what chance a susceptible
has to become infected in a course of an epidemic. The severity of an
epidemic or of an outbreak of the specific epidemic of
an infectious disease depends on many factors including the infectivity
probability per contact $r$, the recovery probability $a$ and spatial
distribution of infected and removed at the outbreak of an epidemic.

In order to examine how the average severity of an epidemic of an infectious
disease depends on the barrier permeability we performed a series of
simulations varying $f$. For each $f$ we measured the average severity of an
epidemic as the average over severities of $100$ epidemic outbreaks each
with
the infectivity probability per contact $r=0.3$ and the recovery probability
$a=0.2$ and strating from the initial conditions described as follows. For
each barrier permeability $f$ each simulation has been performed on a
hexagonal lattice with $10^{4}$ nodes, and with $10$ infectives uniformly
and randomly distributed in the circle $A$ of radius $20$ on the lattice at $%
k=0$, and with $16000$ susceptibles uniformly and randomly distributed on
the lattice, from which in the ring surrounding the circle and consisting of
$1000$ susceptibles we ``vaccinated'', at $k=0,$ randomly and independently
a fraction $f$ of them. Results of this LGCA experiment, which show the
average severity of an epidemic of SIR type as a function of barrier
permeability $f$ are presented in Figure \ref{severity}. It is rather
remarkable that as $f$ approaches zero, the average severity of an epidemic
of SIR type comes close to $49\%$, not much different than the corresponding value
 for the ``uniform'' vaccination strategy, when $r=0.3$ and $a=0.2$. This means that
the ``uniform'' vaccination strategy is practically not better than no
vaccination at all in this case. Even quite sparse ``barrier'' is much more
effective than uniform random vaccination.

\section{Conclusions}

We presented a lattice gas cellular automaton for studying spreads of epidemics
of SIR type. We derived an approximate mean-field type description of the
automaton, and discussed differences between the mean-field approximation
and the results of the simulation using LGCA. We also investigated what
effects can have spatial inhomogeneities in the distribution of various
types of populations on the dynamics of the epidemic process. We
demonstrated that the severity of an epidemic can strongly depend of an
initial spatial distribution of vaccinated individuals.

The lattice gas cellular automaton discussed in this work is certainly
simplistic. It does not take into account many complexities which have to be
considered when one attempts to construct a more realistic automaton model
for an epidemic. However, the important feature of our model is explicitness
of mixing and contact processes. Unlike models based on partial differential
equations, our model is individual-based, and the spread of the infection
occurs due to the motion of individuals and their interactions. In such
models it is quite straightforward to introduce different, non-diffusive type
of motion, and investigate the effects of resulting mixing on the dynamics
of the epidemic process. For example, we are currently investigating
periodic motion with some amount of randomness, that might better represent
the behavior of individuals in human populations. Results of such
experiments will be presented elsewhere. Here we would like only to
reiterate that the description of non-diffusive motion with partial
differential equations is usually very difficult or impossible, excluding
trivial cases such as linear transport. Individual-based models are much
more suitable for this purpose, and we hope that they eventually will help
to built epidemic spatial models, tailored for specific diseases, with high
degree of realism.

\textbf{Acknowledgements}

H. Fuk\'s and A. T. Lawniczak acknowledge partial support from the Natural
Science and Engineering Research Council (NSERC) of Canada, The Fields
Institute for Research in Mathematical Sciences, The Mathematics of
Information Technology and Complex Systems (MITACS-NCE), and ``in-kind''
financial support from Nuptek Systems Ltd. The authors thank Bruno Di
Stefano for his helpful comments.

\providecommand{\href}[2]{#2}\begingroup\raggedright\endgroup

\label{colorplate}

\end{document}